\documentclass[aps,prd,twocolumn,groupedaddress,showpacs]{revtex4}
\usepackage{graphicx}
\usepackage{amsmath}

\newcommand{\remove}[1]{}

\def\be{\begin{equation}}
\def\ee{\end{equation}}
\def\ba{\begin{eqnarray}}
\def\ea{\end{eqnarray}}

\begin{document}

\title{Probing Strongly Coupled Chameleons with Slow Neutrons}

\author{Philippe Brax}
\affiliation{Institut de Physique Th\'eorique, CEA, IPhT, CNRS, URA 2306,
  F-91191Gif/Yvette Cedex, France}
\email{philippe.brax@cea.fr}

\author{Guillaume Pignol}
\affiliation{LPSC, Universit\'e Joseph Fourier, CNRS/IN2P3, INPG, Grenoble, France}
\email{guillaume.pignol@lpsc.in2p3.fr}

\author{Damien Roulier}
\affiliation{Universit\'e Joseph Fourier, Grenoble, France}
\affiliation{Institut Laue-Langevin, Grenoble, France}
\email{roulier@ill.fr}

\date{\today}

\begin{abstract}
We consider different methods to probe chameleons with slow neutrons. Chameleon modify the potential of bouncing neutrons over a flat mirror in the terrestrial gravitational field. This induces
a shift in the energy levels of the neutrons which could be detected in current experiments like GRANIT. Chameleons between parallel plates have a field profile which is bubble-like
and which would modify the phase of neutrons in interferometric experiments.
We show that this new method of detection is competitive with the bouncing neutron one, hopefully providing
an efficient probe of chameleons when strongly coupled to matter.
\end{abstract}

\maketitle
%\begin{document}

\section{Introduction}

The accelerated expansion of the Universe has now been firmly established and confirmed by  several cosmological observables \cite{Astier:2012ba}. However the nature of the Dark Energy driving the late time acceleration of the Universe is still a mystery.
In its  present form, the  concordance model  of cosmology includes a  Dark Energy component simply realised as  a cosmological constant.
Dynamical models of Dark Energy use mostly a scalar field, known as quintessence, rolling down along its potential before nearly stopping in the recent past of the Universe \cite{Copeland:2006wr}.
In this case, the energy of the quintessence field is dominated by its potential energy and its effective pressure  becomes almost opposite to its energy density.
This is enough to generate the acceleration of the expansion of the Universe.
Unfortunately, this comes at a price which is the existence of a long range scalar interaction which could upset the tests of gravity in the solar system, if the quintessence field is coupled to standard model particles.
Successful models of screened modified gravity have been introduced to alleviate this problem \cite{Khoury:2010xi}. Indeed, when coupled to matter, the new fifth force mediated by the quintessence field becomes screened in dense environments preventing its potential detection in the very
stringent laboratory tests of the existence of fifth forces.
One particularly conspicuous class of quintessence  with a screening mechanism  is the chameleon model \cite{KhouryWeltman}.

In this context, slow neutron experiments where neutrons are produced non-relativistically offer an important possibility to test chameleon models. Indeed it turns out that such neutrons are not screened and therefore feel the full strength of the chameleonic interaction. In the case of bouncing neutrons over a flat mirror subject to the terrestrial gravitational field, the chameleons perturb the Newtonian potential and change the energy levels of the neutron. This could  potentially be detected by the GRANIT experiment for large enough values of the coupling of chameleons to matter. A similar sensitivity to the presence of chameleons can be achieved using neutron interferometry where slow neutrons traverse a chamber where the chameleon profile is bubble-like. This disturbs the interference patterns and could therefore give a clear signature of the existence of chameleons.
As both setups are sensitive to relatively large values of the coupling to matter,
we investigate the physics of chameleons in a gas further and show that above a certain density dependent coupling,
the chameleon field in a gas cannot be taken to be a homogeneous constant but develops bubbles between the nuclei. When bubbles form, the sensitivity of the interferometry experiment decreases drastically.

This article is organized as follows,
in section 2 we recall details of chameleon models.
In section 3 we use a numerical integration of Schr\" odinger's equation in the presence of chameleons to evaluate the sensitivity of the quantum states of bouncing neutrons to the chameleonic interaction.
In section 4 we propose to use neutron interferometry as a novel way of detecting chameleons.
In section 5 we study the transition from a homogeneous description of the chameleon field in a gas to a bubble-like situation where the heterogeneities in the gas are seen by the chameleon. We then apply this result to neutron interferometry.
Finally we conclude in section 6.
\section{Chameleons}

Quintessence models accounting for the recent acceleration of the Universe suffer from one major problem as the field generating the acceleration has a very low mass.
This leads to the existence of a fifth force when matter couples to dark energy. Solar system tests of gravity being very stringent, screening mechanisms of the fifth force have been devised.
In that respect, chameleons have been introduced to model the acceleration of the expansion of the Universe \cite{KhouryWeltman,Brax:2004qh} using a scalar field (called the chameleon) whose fifth force is screened very efficiently in the presence of dense matter.
The chameleon dynamics are governed by a  potential $V(\varphi)$ which depends on a single scale $\Lambda$
\begin{equation}
V(\varphi)= \Lambda^4 f(\varphi/\Lambda)
\end{equation}
where $\Lambda$ is determined by the present value of the dark energy, $\Lambda^4= 3 \Omega_{\Lambda 0} H_0^2 M_{\rm Pl}^2 \approx 2.4 \times 10^{-12}$~GeV,
where $H_0$ is the Hubble rate now and $M_{\rm Pl} = 2.44 \times 10^{18}$~GeV is the reduced Planck mass.
When the field is large enough $\varphi\gg \Lambda$, the function converges to one, i.e. $f\to 1$, which can be realised within a large class of runaway potentials.
Moreover, $f$ is assumed to be monotonic (decreasing) and convex guaranteeing that the second derivative of $V$ is positive, i.e. that the mass of the scalar field (in the absence of matter) is positive.
One can choose for instance a Ratra-Peebles model \cite{Brax:2004qh}
\begin{equation}
\label{RatraPeebles}
V(\varphi) = \Lambda^4 + \frac{\Lambda^{4+n}}{\varphi^n}
\end{equation}
where $n>0$ is the Ratra-Peebles index.
For such a model, dark energy is realised when $\varphi \gg \Lambda$ in the absence of coupling to matter and
the equation of state can be close to $w = -1$. The mass of the scalar field is then always less than $\Lambda$ implying that the range of the
scalar interaction is always larger than one millimetre (and very often much larger).
Hence this model of dark energy leads to the existence of a long range scalar force which could be detected in laboratory experiments or solar system tests of gravity, and therefore has to be screened.
However the presence of matter has a direct effect on the potential which becomes the effective potential
\begin{equation}
V_{\rm eff}(\varphi)= V(\varphi) +\frac{\beta}{M_{\rm Pl}} \varphi \rho,
\end{equation}
where $\rho$ is the mass density of matter and $\beta$ is the dimensionless coupling constant of chameleons with matter.
This effective potential is drastically different from $V(\varphi)$ as it possesses a density-dependent minimum $\varphi (\rho)$ with a mass $M(\rho)$ which increases with the density of matter.
More precisely we have that
\be
\varphi(\rho) = \Lambda \left( \frac{n \Lambda^3 M_{\rm Pl}}{\beta \rho} \right)^{1/(n+1)}
\ee
and the mass is given by the curvature of the effective potential at the minimum
\be
\label{range}
M(\rho)^2 = V_{\rm eff}''(\varphi(\rho)) = n(n+1) \frac{\Lambda^{n+4}}{\varphi^{n+2}(\rho)}
\ee
In the following we will consider chameleons in a gas, and we will see that treating the gas density as homogeneous is only an approximation valid when the coupling $\beta$ and the matter density $\rho$ are not too high.

In a dense environment, the mass increases with the matter density and can become very large.
This explains why chameleons cannot be seen in the solar system. Indeed inside large (and screened) objects such as the sun, the field generated by an infinitesimal element is Yukawa-suppressed and does not reach the outer region of the compact body.
Only a thin shell generates any field, which is therefore heavily depleted outside, leading to a negligible deviation from Newton's law.
It turns out that $\Lambda_0$ in the Ratra-Peebles potential $\Lambda_0^{4+n}/\varphi^n$ such that all the gravitational tests are evaded must be $\Lambda_0\le \Lambda$ \cite{KhouryWeltman}.
Hence chameleons can both generate the acceleration of the expansion of the Universe and satisfy the gravity tests of Newton's law with a single scale $\Lambda$.

\section{Quantum States of Bouncing Neutrons}

Ultracold neutrons bouncing over a mirror show a quantum behaviour when the bouncing height is about $10 \ \mu$m.
Namely, the energy of the vertical motion is quantized as for any quantum particle in a potential well.
Experiments are being set up to measure precisely the discrete energy levels of the bouncing neutrons.
We have already shown that strongly coupled chameleons could have an influence on the energy levels \cite{BraxPignol}, taking benefit of the fact that neutrons are not subject to the chameleon screening mechanism.
Using earlier experiments we have set a limit on the chameleon coupling to matter of $\beta < 10^{11}$, depicted by the blue line in fig. \ref{exclusion}.
Recently the QBounce collaboration has reported a measurement of the resonant transitions between low lying quantum states in agreement with the standard theory,
from which they derived the limit $\beta < 5 \times 10^9$ \cite{Jenke}.
This limit is also reported in fig. \ref{exclusion}.

In \cite{BraxPignol} we have treated the effect of the chameleon field on the energy spectrum at first order in perturbation theory.
We will confirm  the validity of this previous calculation with an exact treatment.
In the absence of the chameleon, the bouncing neutron potential is \mbox{$\Phi(z)=mgz$}  with $m$ the neutron mass and \mbox{$g=9.806$~m.s$^{-2}$} is the acceleration of gravity in Grenoble.
In the presence of the chameleon interaction, the chameleon field acquires a universal profile independent of $\beta$ above the mirror.
The interaction potential is then modified:
\begin{equation}
\Phi(z)=m gz+\beta V_n\left(\displaystyle\Lambda z\right)^{\alpha_n}
\end{equation}
with \mbox{$V_n=\displaystyle ({m}/{M_{Pl}})\Lambda\left(({2+n})/{\sqrt{2}}\right)^{{2}/({2+n})}$} and \mbox{$\alpha_n=\displaystyle {2}/({2+n})$}.
The stationary Schr\"odinger equation for the vertical motion (along $z$) of the bouncing neutron becomes
\begin{equation}
-\displaystyle\frac{\hbar^2}{2m}\frac{d^2}{dz^2}\psi+\Phi(z)\psi(z)=E\psi(z)
\end{equation}
where $\psi$ is the wave function (with $\psi(0)=0$ on the mirror) corresponding to the quantum state of energy $E$.
Without the chameleon, the unperturbed wave functions of the neutron in the terrestrial gravitational field are given by the Airy functions
\begin{equation}
\label{eq:airy}
\psi_k(z)=c_k {\rm Ai} \left(\displaystyle\frac{z}{z_0} - \epsilon_k \right)
\end{equation}
and $ E_k=E_0\epsilon_k$
where $c_k$ is a normalization constant, \mbox{$E_0 = mgz_0 = 0.6$~peV} and \mbox{$z_0=\left(\displaystyle\frac{\hbar^2}{2m^2g}\right)^{\frac{1}{3}}=5.87$ $\mu$m}.
The values of $\epsilon_k$ are the zeros of the Airy function \mbox{$\{ \epsilon_k \}_{k = 1, 2 \cdots} = \{ 2.338 , 4.088 , 5.521 , 6.787 , 7.944 , 9.023 \cdots \}$}.

Treating the chameleon potentiel as a perturbation, the shifted energy levels are given by
\begin{equation}
\delta E_k= \beta V_n \langle \psi_k\vert \left(\displaystyle\Lambda z\right)^{\alpha_n}\vert \psi_k \rangle
\end{equation}
where $\vert \psi_k \rangle$ is the k-th level wave function.
Thus the shift on an energy level at first order in perturbation theory can be obtained using the matrix elements
\begin{equation}
O_{k}(\alpha)= \langle \psi_k | \left(\displaystyle\frac{z}{z_0}\right)^\alpha | \psi_k \rangle.
\end{equation}
These overlap functions are tabulated in table I.

\begin{table}
\label{overla}
\caption{\label{tab:overlap}Overlap functions}
\begin{ruledtabular}
\begin{tabular}{c c c c c c c c c}
 n & 1 & 2 & 3 & 4 & 5 & 6 & 7 & 8\\
$\alpha_n$ & $\frac{2}{3}$ & $\frac{1}{2}$ & $\frac{2}{5}$ & $\frac{1}{3}$ & $\frac{2}{7}$ & $\frac{1}{4}$ & $\frac{2}{9}$ & $\frac{1}{5}$\\
 \hline
$O_1(\alpha_n)$ & 1.31 &1.22 & 1.16 & 1.13 &1.11 & 1.09 & 1.08 & 1.07\\
$O_2(\alpha_n)$ & 1.89 &1.59 & 1.44 & 1.35 &1.29 & 1.25 & 1.22 & 1.19\\
$O_3(\alpha_n)$ & 2.31 &1.85 & 1.62 & 1.49 &1.40 & 1.34 & 1.30 & 1.26\\
 \end{tabular}
 \end{ruledtabular}
 \end{table}

\begin{figure}

\includegraphics[width=0.95\linewidth]{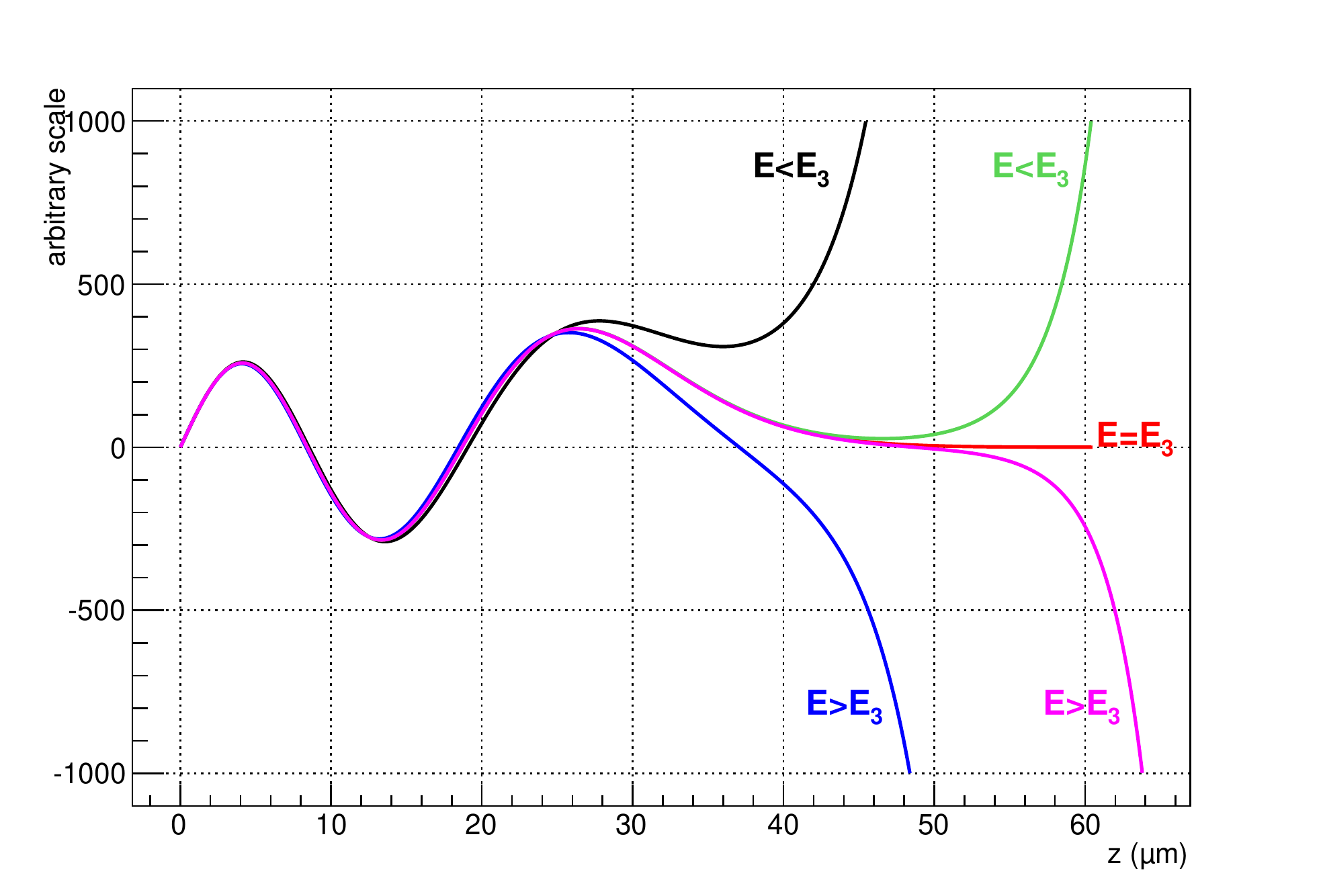}
\caption{\label{fig:divergence}Calculated wavefunction for energies close to $E_3$, for the gravitational field only}
\end{figure}

Now let us go beyond perturbation theory and compute the energy eigenvalues exactly in the presence of the chameleon field.
The numerical resolution of the Schr\"odinger equation (for $z=0$ to $\infty$) for any energy can be done numerically (here with Numerov's method \cite{Landau}) starting from $\psi(0)=0$ and $\psi'(0)=1$ . For the energy corresponding to the quantum level $k$, the wavefunction converges toward zero after the  $k$ extrema (see fig. \ref{fig:divergence}).
The method to find the energy levels $E^*_k$ is therefore the following :
\begin{itemize}
\item If the wavefuncton has less than $k$ extrema, it cannot be the solution of the equation for $E^*_k$.
\item If the wavefunction has at least $k$ extrema, and if after them it has another extremum without changing sign, then $E<E^*_k$ and the wavefunction  diverges.
\item If the wavefunction has $k$ extrema, and if it converges towards zero, then $E=E^*_k$.
\item If the wavefunction has at least $k$ extrema, and if after them it changes sign, then $E>E^*_k$ and the wavefunction  either diverges, or has another extremum .
\end{itemize}
Applying this dichotomic method, we can stop the numerical calculation very quickly and converge towards the value of $E^*_k$.
This allows us to find the energy levels $E^*_k=E_0\epsilon^*_k$ by dichotomy for the chameleon correction to the  potential.

\begin{figure}
\includegraphics[width=0.95\linewidth]{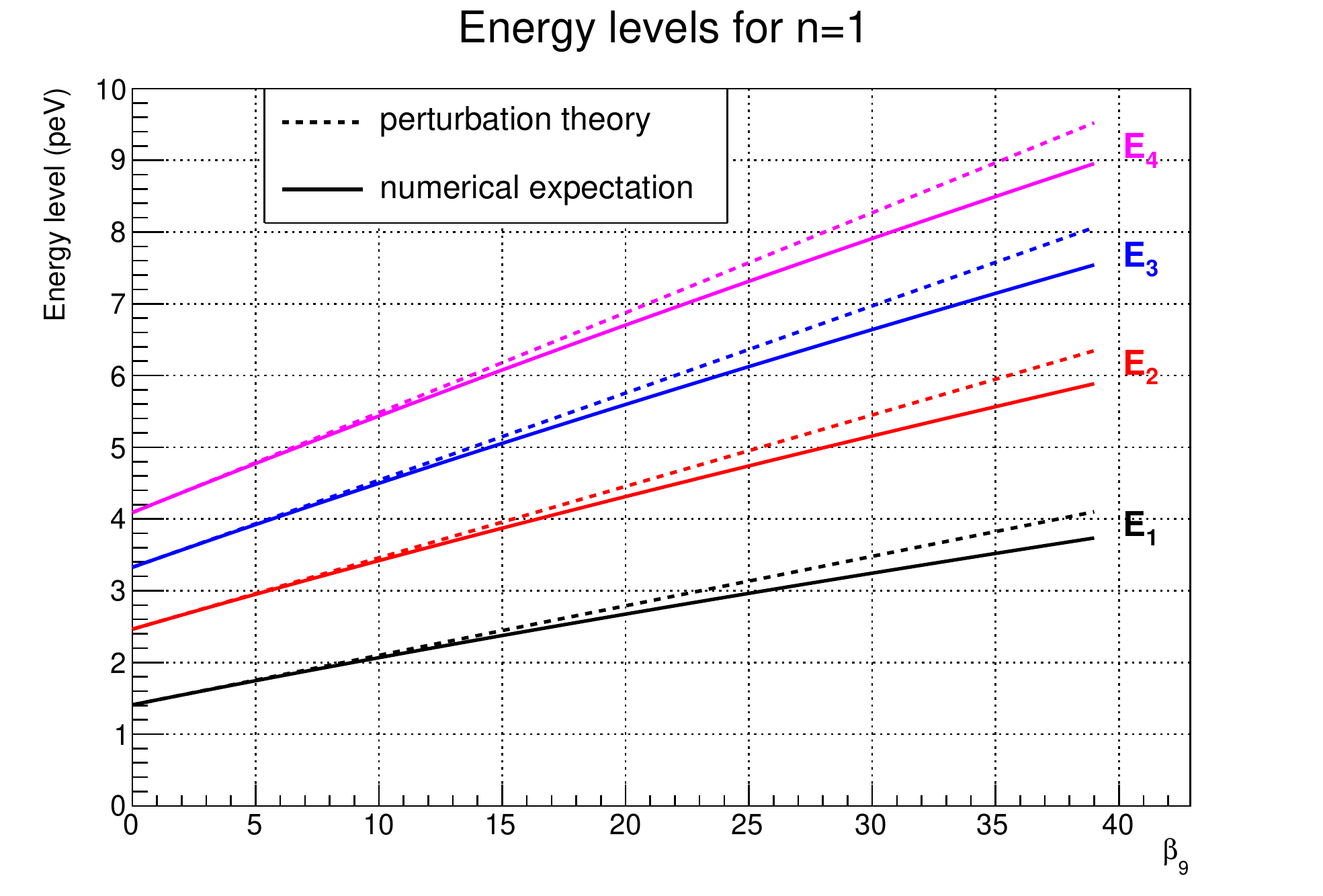}
\includegraphics[width=0.95\linewidth]{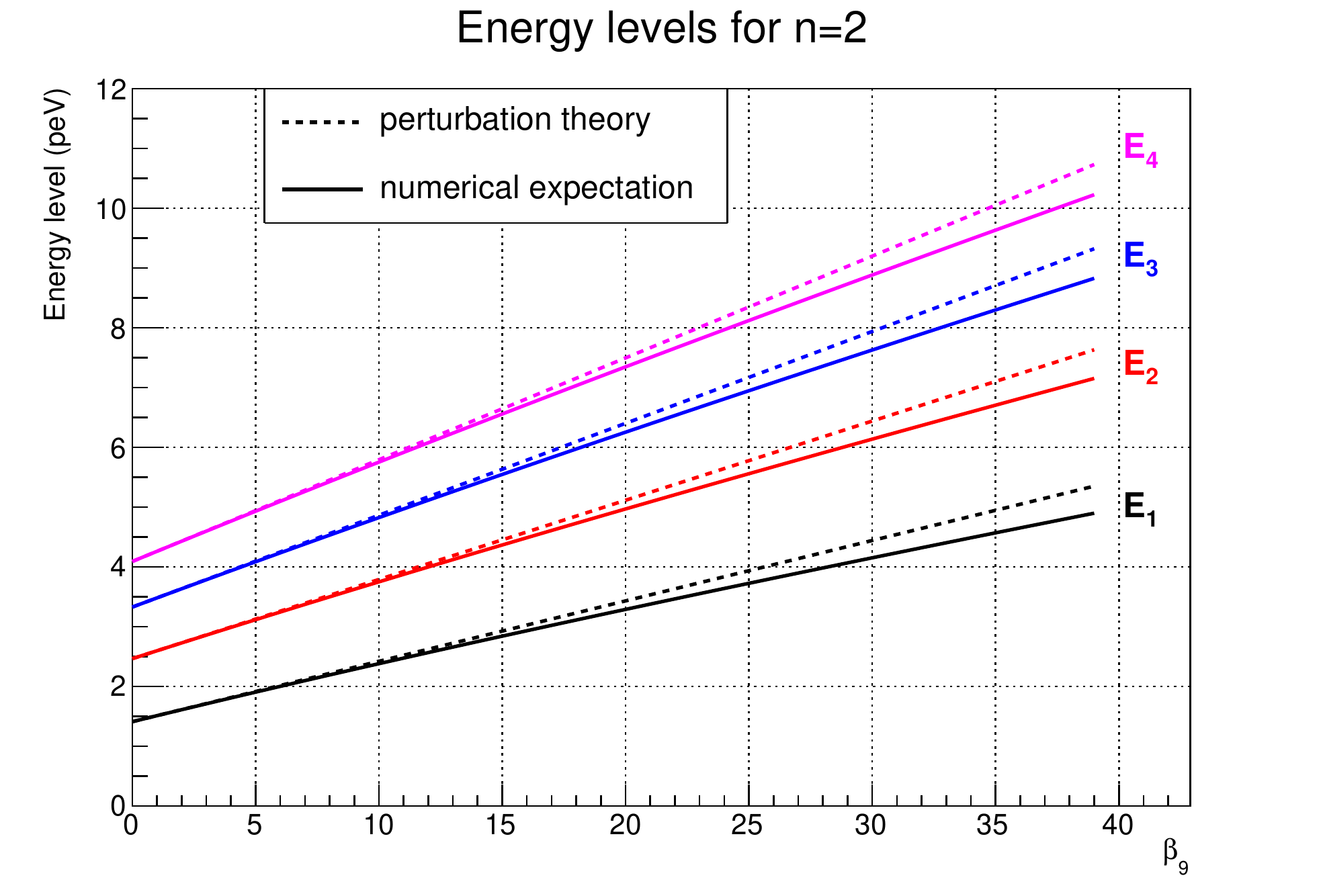}
\includegraphics[width=0.95\linewidth]{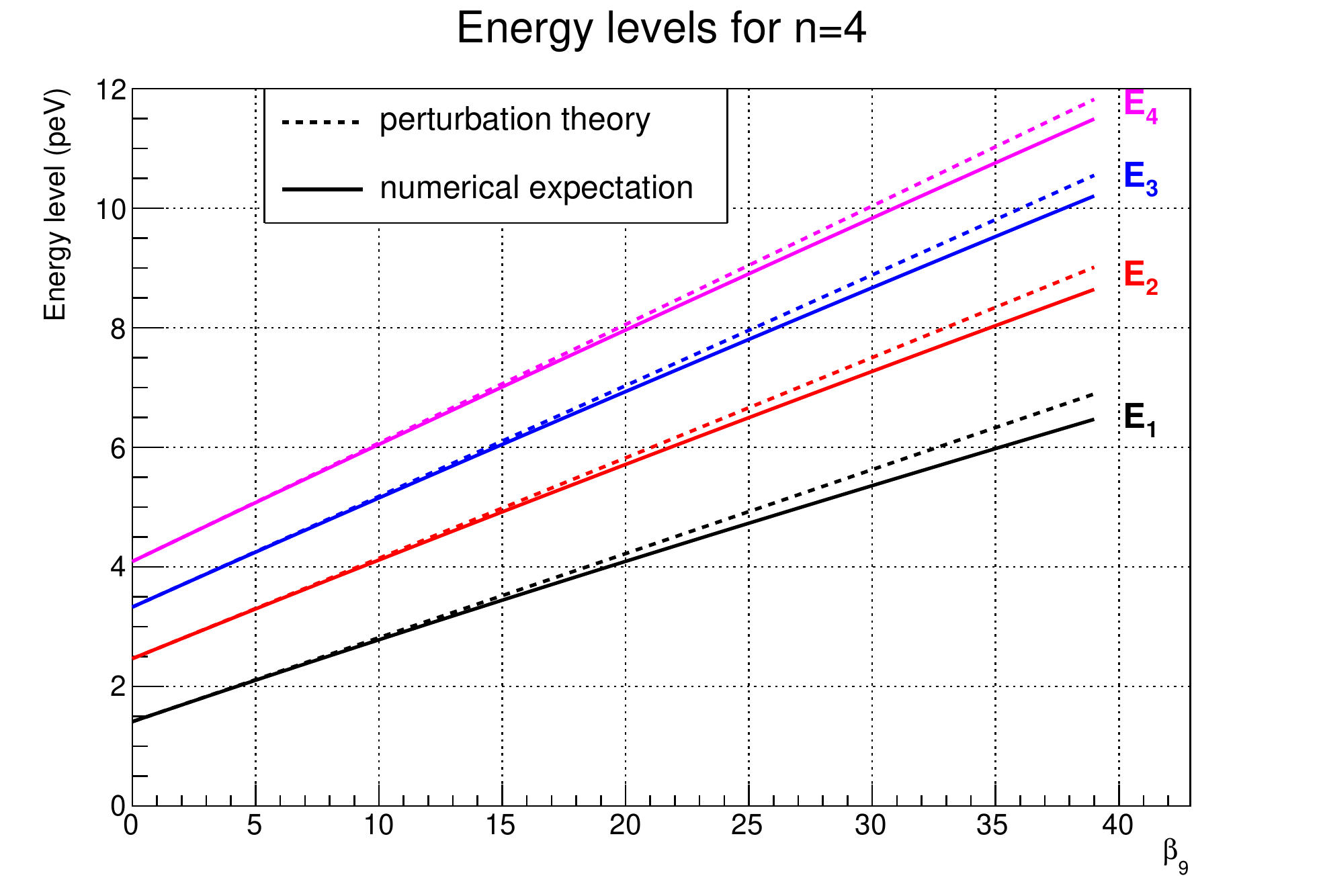}
\caption{\label{fig:shift}
Energy spectrum of the neutron quantum bouncer as a function of the chameleon coupling $\beta_9 = 10^{-9} \times \beta$ calculated at first order in perturbation theory (dashed line) and with numerical precedure described in the text (bold lines).
}
\end{figure}

We have solved numerically Schr\"odinger's equation for \mbox{$\Phi(z)=mgz$} and obtained the numerical solutions for the zeros $\epsilon_k$ of the Airy function $Ai$ using Numerov's method. An integration step of $0.01$~$\mu$m and a $z$-range of $100$~$\mu$m gives a $10^{-5}$ precision on $\epsilon_k$.
We have compared the energy shifts predicted by perturbation theory to the numerical solution of the 1D-Schr\"odinger equation (see Fig. \ref{fig:shift}), for the first four energy levels of the bouncing neutron.

\begin{figure}

\includegraphics[scale=0.35]{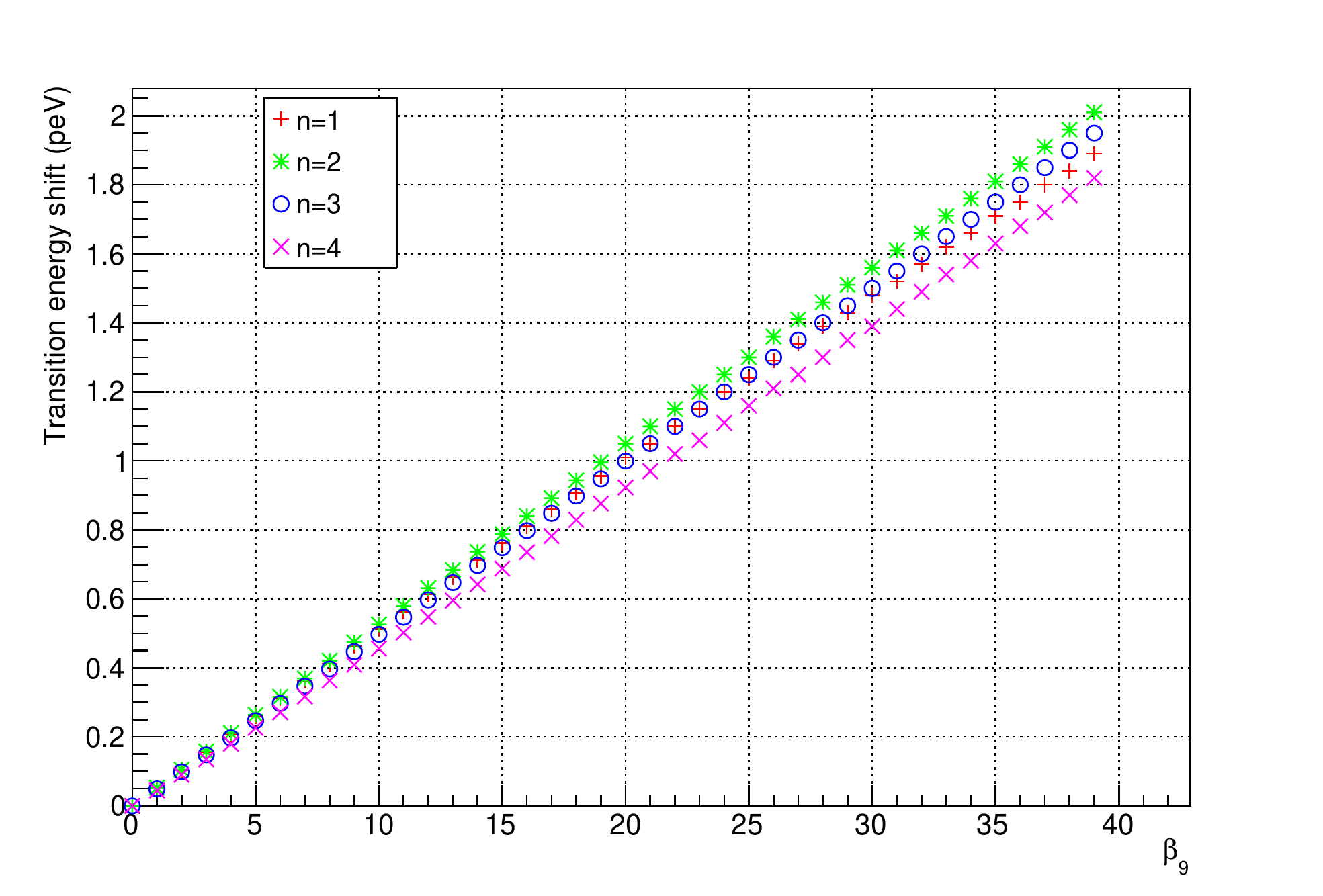}
\caption{\label{fig:shift_transition}Calculated energy shift due to chameleon in the transition $3\rightarrow 1$}
\end{figure}

The validity of the perturbation analysis breaks down  for $\beta>10^{10}$ and full numerical results are compulsory.
For $\beta < 10^{10}$ the accuracy of the perturbation theory is good enough to estimate the limits of chameleon couplings, which validates the previous analysis \cite{BraxPignol}.
Using our numerical results, we are able to calculate the shift induced by the chameleon potential on the energy of transition (see fig. \ref{fig:shift_transition}).
In GRANIT, the gap between two energy states such as $k=3$ and $k=1$ will be precisely measured with an estimated accuracy of $0.01$ peV compared to the nominal energy $E_3-E_1=1.91$~peV \cite{GRANIT}.
By requiring that the chameleonic shift does not exceed the expected sensitivity, we get a bound on the coupling $\beta$ which depends on $n$.
Fig. \ref{exclusion} plots the part of the parameter space which will be covered by the GRANIT experiment.
Coupling strengths in the range $\beta \approx 10^8$ are within reach of the expected sensitivity of GRANIT.
The precision of the quantum levels is set by the observation time according to Heisenberg's relation.
Ultimately, the precision will be limited by the neutron lifetime and the associated relative precision is about $10^{-7}$.
In this case, for an experiment storing neutrons in quantum states for several minutes, coupling strengths of $\beta \approx 10^3$ could be detected.

\section{Neutron interferometry}

\begin{figure}
\centering
\includegraphics[width=0.95\linewidth]{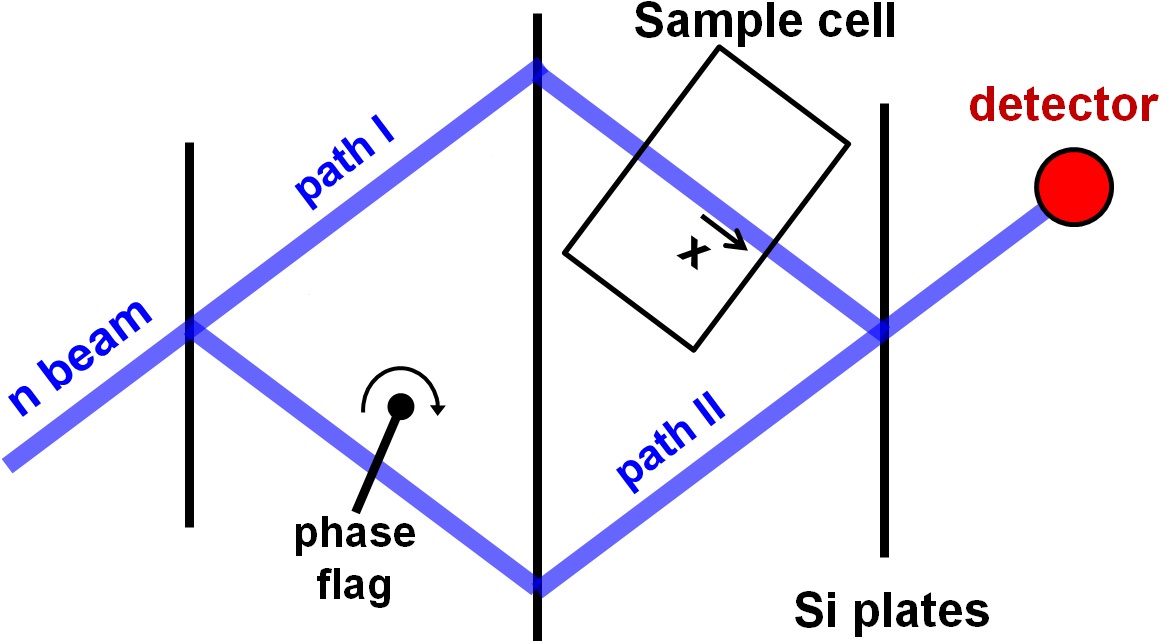}
\caption{Sketch of the neutron interferometer setup.
}
\label{interferometer}
\end{figure}

Neutron interferometry provides a second method to probe the chameleon field.
It has been recently  proposed  to develop a Loyd's type interferometer with very cold neutrons
which can be sensitive to the chameleon potential gradient in the vicinity of a mirror \cite{Pokotilovski}.
Such a method could in principle be sensitive to chameleon couplings down to $\beta = 10^7$, but the cold neutron interferometer technique has yet to be developed.
As an alternative, we propose to use LLL interferometers with slow neutrons that have been  operated routinely for decades in several neutron facilities.

The sketch of a typical setup is depicted in fig. \ref{interferometer}.
A monochromatic neutron beam with a wavenumber of $k = 23$~nm$^{-1}$ is split into two coherent beams using a mono-crystal silicon plate.
Then part of these two beams are recombined using two additional similar parallel plates.
The neutron detector measures the flux resulting from the interference of neutrons going through path I and path II.
A phase flag (usually an aluminium plate with variable angle) is introduced in path II to record the interference pattern:
the neutron flux measured by the detector is an oscillating function of the phase flag angle.
A sample is introduced in  path I.
Measuring the interference pattern with and without the sample, one can extract the phase shift of the sample.

For the purpose of detecting chameleons, the sample will consist of a cell with parallel plates normal to the neutron beam as shown in fig \ref{interferometer}.
We will assume that the transverse dimension of the cell is infinite and denote the distance between the plates by $2R$.
For numerical studies we will set $2 R = 1$~cm as in the experiment performed at NIST \cite{NIST}.
When the cell is filled by no gas, i.e. in vacuum, a chameleon bubble-like profile $\varphi(x), -R < x < R$, will appear in the cell, inducing the potential $\beta m/M_{\rm Pl} \ \varphi(x)$ for the neutrons.
One can then show (see e.g. \cite{Rauch}) that the phase shift due to the chameleon bubble is given by
\begin{equation}
\label{phaseShift}
\delta \phi = \frac{m}{k \hbar^2} \int_{-R}^{R} \beta \frac{m}{M_{\rm Pl}} \varphi(x) dx.
\end{equation}

We will calculate the chameleon bubble integral $\int \varphi(x) dx$ and show that it can be suppressed when introducing a  gas at moderate pressure in the sample cell.
The chameleon profile $\varphi(x)$ in the sample cell satisfies the 1D chameleon equation
\begin{equation}
\label{KleinGordon}
\frac{d^2 \varphi}{dx^2} = V'(\varphi) + \frac{\beta}{M_{\rm Pl}} \rho
\end{equation}
where $\rho$ is the mass density of the gas inside the sample cell.
We will assume the boundary condition $\varphi(-R) = \varphi(R) = 0$,
which proves to be valid when $\varphi_c \ll \varphi_0$ where $\varphi_c$ is the field value inside the plate bulk and $\varphi_0 = \varphi(0)$ is the maximum of the field in the sample cell.

\begin{figure}
\centering
\includegraphics[width=0.9\linewidth,angle=90]{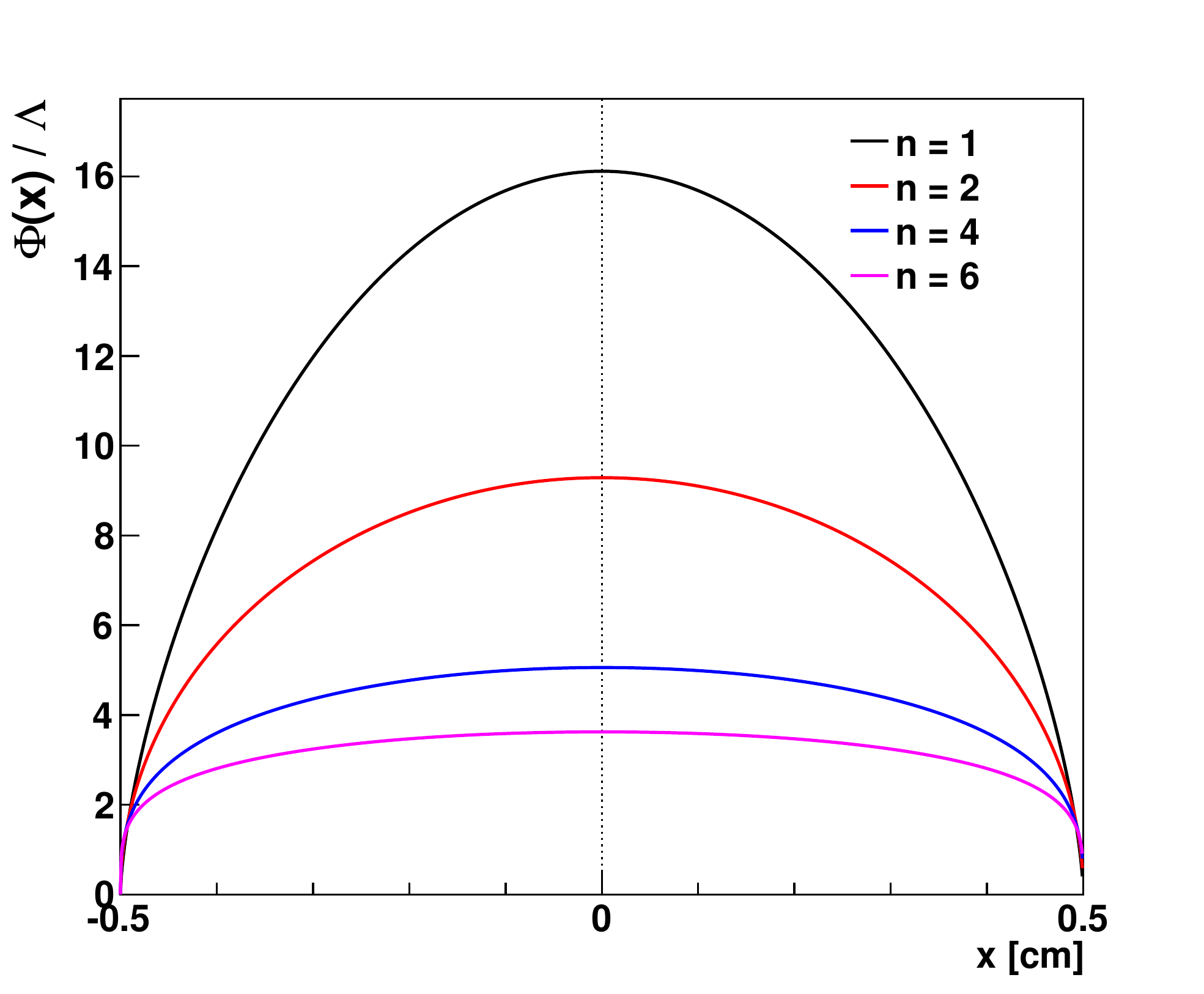}
\caption{
Chameleon bubble profile in vacuum between two plates separated by a distance of 1~cm.
}
\label{bubble}
\end{figure}

The problem in the case of perfect vacuum in the cell $\rho = 0$ has been addressed in \cite{Ivanov}.
They found an analytical form for the chameleon field profile:
\begin{equation}
\label{IvanovBubble}
\varphi(x) = \Lambda ( R \Lambda )^{2/n+2} \ \left( \frac{n+2}{2 \sqrt{2}} \left[ 1 - (z/R)^2 \right]\right)^{2/n+2}
\end{equation}
which is exact for $n=2$ and valid with an accuracy better than $4 \, \%$ for $n>2$.
We plot the vacuum solution (\ref{IvanovBubble}) in fig. \ref{bubble} in the case $2R = 1$~cm.
It is apparent that the chameleon field forms bubbles in vacuum. We have also represented the field profile in 2D obtained numerically in figure 7 when the transverse dimension of the chamber is finite. We find that bubbles still form in this geometry.

\begin{figure}
\centering
\includegraphics[width=0.87\linewidth,angle=90]{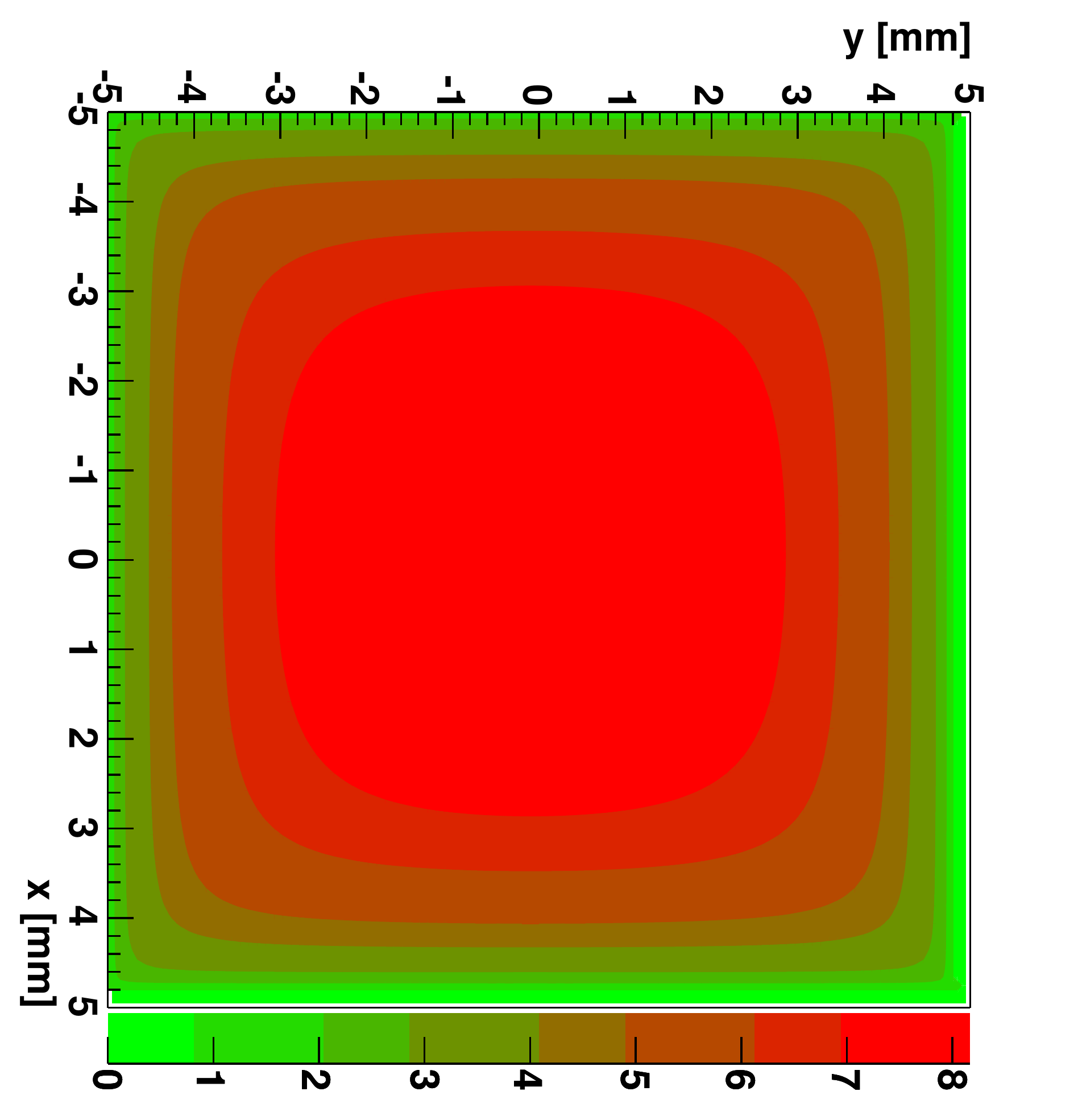}
\caption{The chameleon profile $\varphi / \Lambda$ in 2D in a square box calculated for $n=2$.}

\label{exclusion}
\end{figure}

Let us now elaborate in  the case $\rho > 0$.
First we transform eq. (\ref{KleinGordon}) into
\begin{equation}
\frac{dz}{d \varphi} = \frac{1}{\sqrt{2}} \left( V(\varphi) - V(\varphi_0) + \frac{\beta \rho}{M_{\rm Pl}} (\varphi - \varphi_0) \right)^{-1/2}.
\end{equation}
The maximum $\varphi_0$ can be calculated by the implicit relation
\begin{equation}
\label{toFindPhi0}
\int_0^{\varphi_0} \frac{dz}{d \varphi} d \varphi = R.
\end{equation}
Then one can evaluate the bubble integral
\begin{eqnarray}
\label{toCalculateBubbleIntegral}
\int_{-R}^R \varphi(x) dx & = & 2 \int_0^{\varphi_0} \varphi \frac{dz}{d\varphi} d \varphi \\
\nonumber
& = & \sqrt{2} \int_0^{\varphi_0} \frac{\varphi d\varphi}{ \sqrt{ V(\varphi) - V(\varphi_0) + \frac{\beta \rho}{M_{\rm Pl}} (\varphi - \varphi_0) }}
 \end{eqnarray}
Let us now apply  these results to the case of the Ratra-Peebles potential for chameleons (\ref{RatraPeebles}).
We define
\begin{eqnarray}
K_n(\alpha) & = & \int_0^1 \frac{u^{1+n/2} \ du}{\sqrt{1-u^n + n \alpha \, u^n (u-1)}} \\
J_n(\alpha) & = & \int_0^1 \frac{u^{n/2} \ du}{\sqrt{1-u^n + n \alpha \, u^n (u-1)}}.
\end{eqnarray}
Furthermore with $y_0 = \varphi_0 / \Lambda$,
eq. (\ref{toFindPhi0}) becomes
\begin{equation}
\label{implicit_y0}
\sqrt{2} \, R \, \Lambda = y_0^{n/2 + 1} \, J_n \left( \frac{\beta \rho}{M_{\rm Pl} \Lambda^3} \frac{y_0^{n+1}}{n} \right).
\end{equation}
This is an implicit equation  determining the bubble maximum $y_0$.
Then the bubble integral (\ref{toCalculateBubbleIntegral}) can be expressed as a function of $y_0$ as
\begin{equation}
\label{implicit_bubbleIntegral}
\int_{-R}^R \varphi(x) dx \approx \sqrt{2} \, y_0^{n/2+2} K_n \left( \frac{\beta \rho}{M_{\rm Pl} \Lambda^3} \frac{y_0^{n+1}}{n} \right)
\end{equation}
Two limiting cases can be considered: the low pressure and high pressure cases.

With perfect vacuum in the cell $\rho = 0$, eq. (\ref{implicit_y0}) is not implicit anymore and the bubble integral (\ref{implicit_bubbleIntegral}) can be expressed exactly as
\begin{equation}
\label{vacuum_bubbleIntegral}
\int_{-R}^R \varphi(x) dx = \sqrt{2} \left( \frac{\sqrt{2} R \Lambda}{J_n(0)}\right)^{\frac{n+4}{n+2}} \, K_n(0).
\end{equation}

In the high pressure case, when the range of the chameleon (\ref{range}) becomes smaller than the cell size $R$,
the field $\varphi$ settles at the minimum of the potential $V_{\rm eff}'(\varphi) = 0$.
Indeed, by inspection, eq. (\ref{implicit_y0}) shows that the argument of the $J_n$ function  approaches unity for high $\rho$.
Then the bubble integral (\ref{implicit_bubbleIntegral}) at high pressure becomes
\begin{equation}
\label{highpressure_bubbleIntegral}
\int_{-R}^R \varphi(x) dx \approx 2 R \Lambda \left( \frac{n M_{\rm Pl} \Lambda^3}{\beta \rho} \right)^{1/n+1}.
\end{equation}
This result is valid as long as the gas does not become a heterogeneous medium for chameleons, which happens at large coupling and/or large pressure. We will reassess this calculation for the heterogeneous case at very large coupling in the following section.

\begin{figure}
\centering
\includegraphics[width=0.9\linewidth,angle=90]{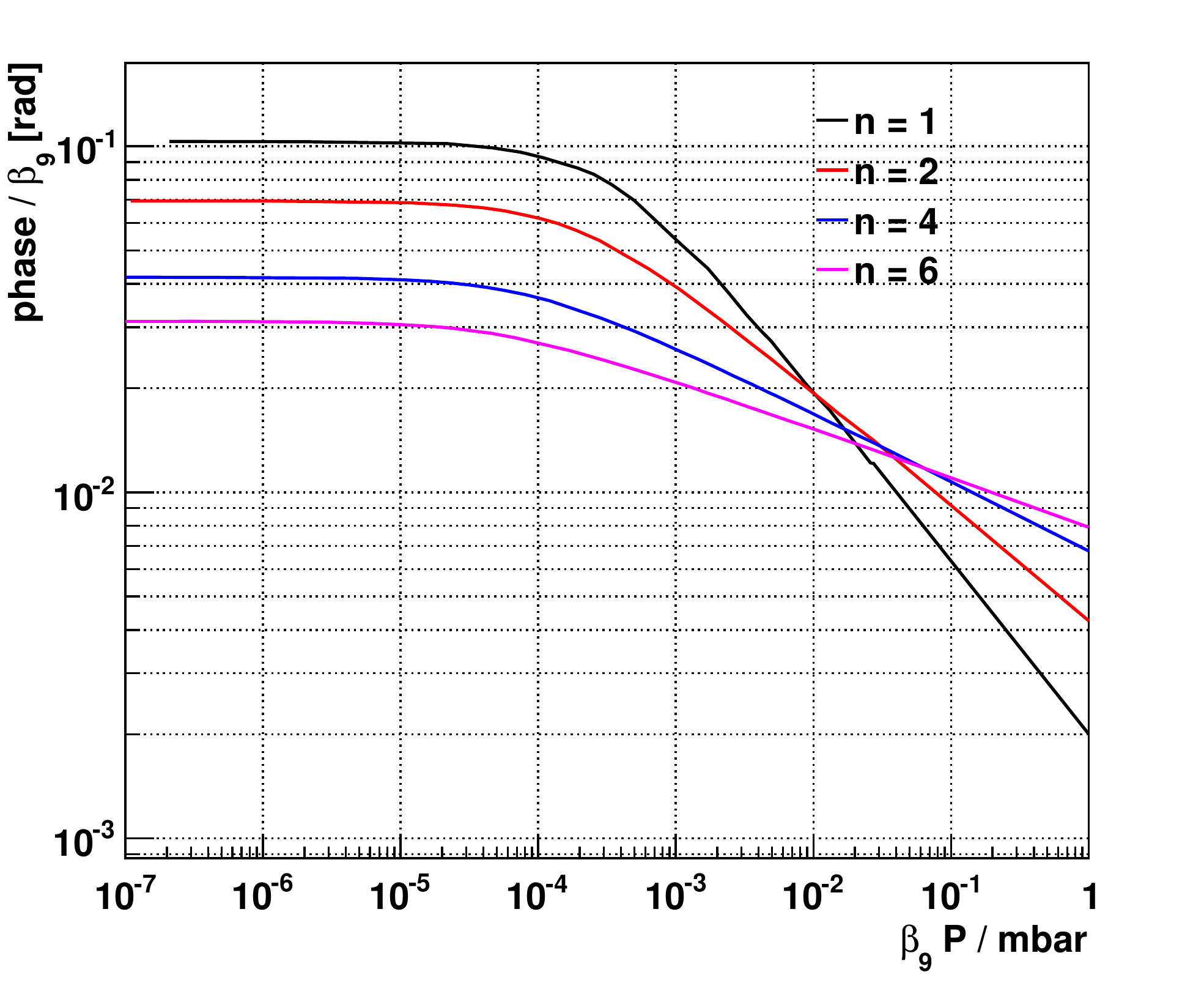}
\caption{Phase $\delta \phi$ due to the chameleon bubble as a function of the helium pressure in the cell with dimension $2R = 1$~cm, with $\beta = \beta_9 \times 10^9$.
}
\label{phaseVSpressure}
\end{figure}

To calculate the phase shift (\ref{phaseShift}) due to the chameleon bubble in the intermediate pressure regime, when the chamelon range is similar with the size of the cell,
we have solved numerically eq. (\ref{implicit_y0}) and (\ref{implicit_bubbleIntegral}).
Assuming helium in the cell we find numerically
\begin{equation}
\frac{\beta \rho}{M_{\rm Pl} \Lambda^3} = 23 \, \beta_9 \, \frac{P}{1 \ {\rm mbar}}
\end{equation}
where $P$ is the pressure of the gas (helium) and $\beta_9 = 10^{-9} \ \beta$.
The result is shown in fig. \ref{phaseVSpressure}.
Since the typical sensitivity of neutron interferometers is $\delta \phi = 1 \deg = 17$~mrad,
this technique can probe chameleon couplings in the range $10^8 - 10^9$ depending on the value of the Ratra-Peebles index $n$.
Within this range, it is possible to suppress the chameleon bubble if the cell is filled with helium with a pressure as low as $10^{-1}$~mbar.
This feature makes it possible to switch on and off the chameleon bubble.
Since almost all neutron interferometer experiments are carried out at atmospheric pressure, a dedicated experiment searching for a ``phase shift of the vaccum'' is needed.
The sensitivity of such an experiment is depicted by the green zone labeled ``neutron interferometry'' in fig. \ref{exclusion} assuming a sensitivity of $\delta \phi = 1 \deg$ and a cell size of $2 R = 1$~cm.

\begin{figure}
\centering
\includegraphics[width=0.92\linewidth,angle=90]{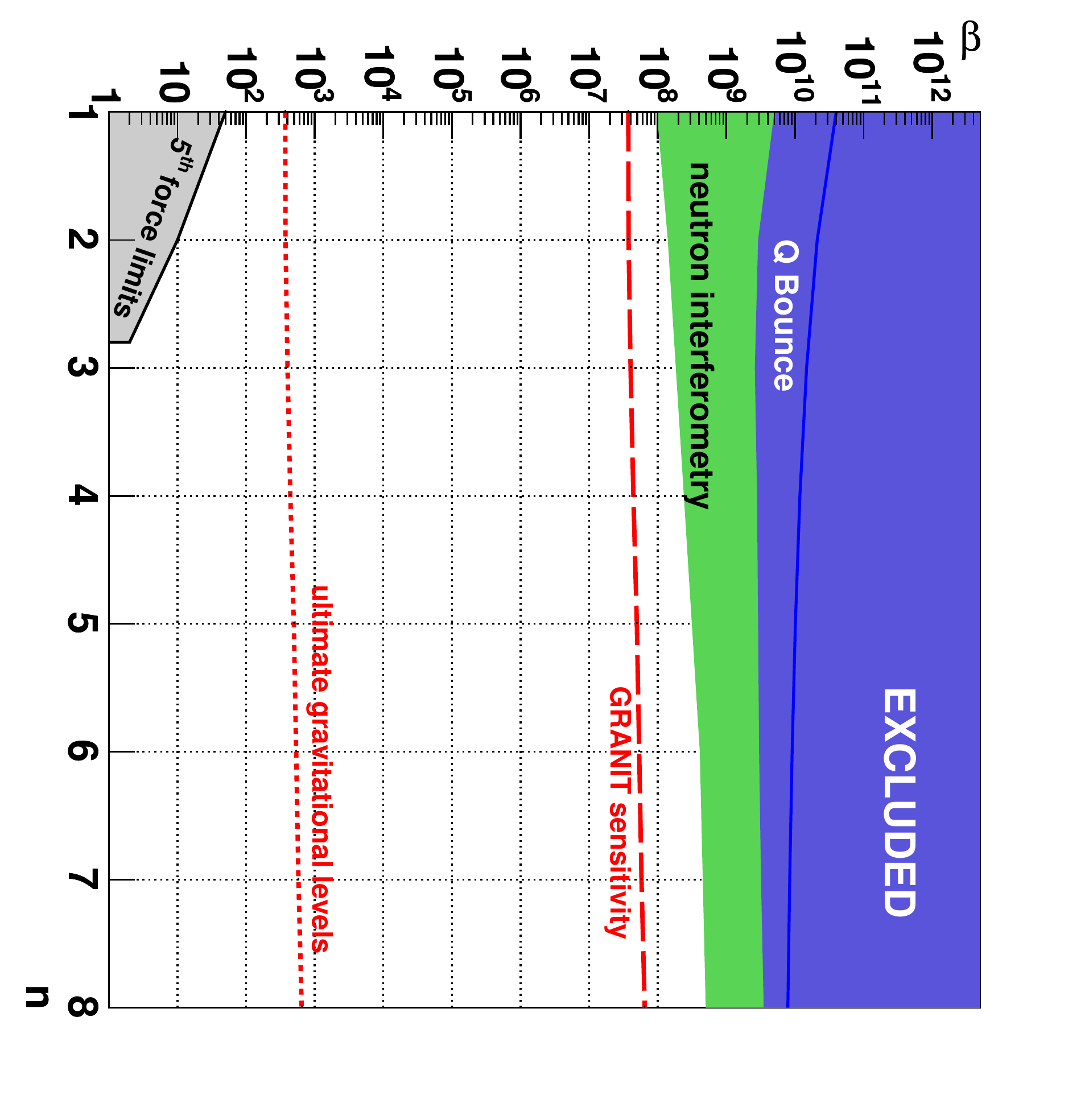}
\caption{The chameleon exclusion plot.
The blue zone is excluded from \cite{Jenke}.
The red lines are sensitivities calculated in \cite{BraxPignol} and confirmed in this work.
The green zone is the potential reach of the neutron interferometry experiment proposed in this work.
}
\label{exclusion}
\end{figure}

\section{Chameleon Bubbles}

When the coupling and/or the density of a gas becomes high enough, the homogeneous approximation which we have used so far, i.e. assuming that the density of matter is supposed to be homogeneous, does not work anymore.
The sub-structure of the gas at the level of each atom must be taken into account. A similar analysis has been be done in \cite{Mota:2006fz}. Here we provide a more thorough analysis of the non-linear phase when the chameleon feels the atomic structure. We then show that
this alters the predictions of neutron interferometry and reduces the sensitivity of such experiments.

\subsection{The homogenous approximation}

The effective potential in the presence of matter is given by
\be
V_{\rm eff}(\varphi)= V(\varphi) +  \beta\frac{\varphi}{M_{\rm Pl}}  \rho
\ee
When the density can be considered to be homogeneous, the effective potential
 has a minimum determined by the macroscopic density $\bar \rho$
\be
\bar \varphi= \left( \frac{n\Lambda^{4+n} M_{\rm Pl}}{\beta \bar \rho} \right)^{1/(n+1)}
\ee
when $\bar \rho$ is the averaged density of the gas considered as a homogeneous medium.
In the case of a gas of particles, the density is not homogeneous and can be written as
\be
\rho=\sum_i m_{\rm nucl} \delta (x-x_i)
\ee
where we have approximated the density inside the nuclei as a Dirac function. The mass $m_{\rm nucl}$ is the mass of the nuclei.
Upon  averaging over the motion of the atoms we have $<\rho>= \bar\rho$ the macroscopic density of the gas.
The microscopic Klein-Gordon equation is
\be
-\ddot\varphi +\Delta \varphi= V'(\varphi) +\frac{\beta}{M_{\rm Pl}} \rho
\ee
where the density is time dependent due to the motion of the atoms. Expanding around  $\bar \varphi$ the Klein-Gordon equation becomes
\be
-\delta \ddot\varphi +\Delta \delta\varphi= V'(\bar \varphi+\delta\varphi ) +\frac{\beta}{M_{\rm Pl}} \rho
\ee
whose averaged solution is the homogeneous $\bar\varphi$. The profile of $\delta\varphi$ is due to the presence of nuclei where the density is larger than
the homogeneous one obtained by averaging over the motion of the nuclei.

The effective potential can be expanded around the homogeneous field $\bar\varphi$ as
\begin{eqnarray}
V_{\rm eff}(\bar \varphi+\delta\varphi) & = & V_{\rm eff}(\bar \varphi) +\Lambda^4 \left( \frac{\Lambda}{\bar\varphi} \right)^n \sum_{p>1} c_p (\frac{\delta \varphi}{\bar\varphi})^p \\
\nonumber
& & + \frac{\beta}{M_{\rm Pl}} \left( \delta \rho \bar \varphi + \delta \rho \delta \varphi \right)
\end{eqnarray}
where $c_p= (-1)^p \frac{n(n+1)\dots (n+1-p)}{n!}$.
The dominant term in $\delta \varphi/\bar\varphi$ is the mass term for $p=2$ as long as $\delta\varphi \lesssim \bar\varphi$ which guarantees that higher order terms become less and less relevant.
Moreover we shall require that the model is perturbatively valid, i.e. that the self-coupling of the chameleon is small. This is valid when the term in $\delta\varphi^4$ has a small coupling, which occurs when $\bar\varphi\gtrsim \Lambda$.
This correspond to the densities
\be
\rho\lesssim\rho_{\rm pert}
\ee
where
\be
\rho_{\rm pert}=\frac{n\Lambda^3 M_{\rm Pl}}{\beta}.
\ee
In this regime, we can neglect the chameleon self-interactions and safely reduce the dynamics of $\varphi$ to the one of a massive scalar field. As a result,
 the Klein Gordon equation for $\delta \varphi$ reduces to
\be
-\delta\ddot\varphi +\Delta\delta\varphi - M(\bar \rho)^2 \delta\varphi= \beta \frac{\delta \rho}{M_{\rm Pl}}
\ee
where $\delta\rho$ is the deviation from the averaged density.
We can single out one atom and its surrounding cell  where no other atom is present
\be
\rho= m_{\rm nucl} \delta(x-x_j) +\sum_{i\ne j} m_{\rm nucl} \delta (x-x_i)
\ee
and the averaged
\be
<\sum_i m_{\rm nucl} \delta (x-x_i)>= \frac{\bar N-1}{V} m_{\rm nucl}
\ee
where $\bar N$ is the (very large) number of atoms in the gas and $V$ its volume. As $\bar N\gg 1$, this is equivalent to $\bar\rho$ and therefore
in a cell around one particular atom and working in the coordinate frame where the atom is fixed, the Klein-Gordon equation becomes time independent
\be
\Delta\delta\varphi - M(\bar \rho)^2 \delta\varphi= \beta \frac{m_{\rm nucl}}{M_{\rm Pl}}\delta
\ee
In fact, for all the pressure and $\beta$ that we consider, we have $M(\bar \rho) D \ll 1$ where $D$ is the inter atomic distance. This implies that the Klein-Gordon equation becomes
\be
\Delta\delta\varphi= \beta \frac{m_{\rm nucl}}{M_{\rm Pl}}\delta
\ee
This description is valid as long
as the value of $\bar\varphi\gtrsim \vert\delta\varphi\vert $. This constraint is the strongest  at a distance equal to the radius of the nuclei $R_{\rm nuc}$.
The homogeneous approximation is thus valid when
\be
\bar \varphi \gtrsim 2 \beta M_{\rm Pl} \Phi_N
\ee
where $\Phi_N$ is the Newtonian  potential at the surface of the nucleus $\Phi_N= \frac{m_{\rm nucl}}{8\pi M_{\rm Pl}^2 R_{\rm nucl}}$, where $m_{\rm nucl}$ is the mass of the nucleus.
When this is the case the nuclei are not screened and the homogeneous approximation is valid.
When this is not the case anymore, the solution cannot remain homogeneous and chameleon bubbles form as we shall see in the following section.

\begin{figure}
\centering
\includegraphics[width=0.87\linewidth,angle=90]{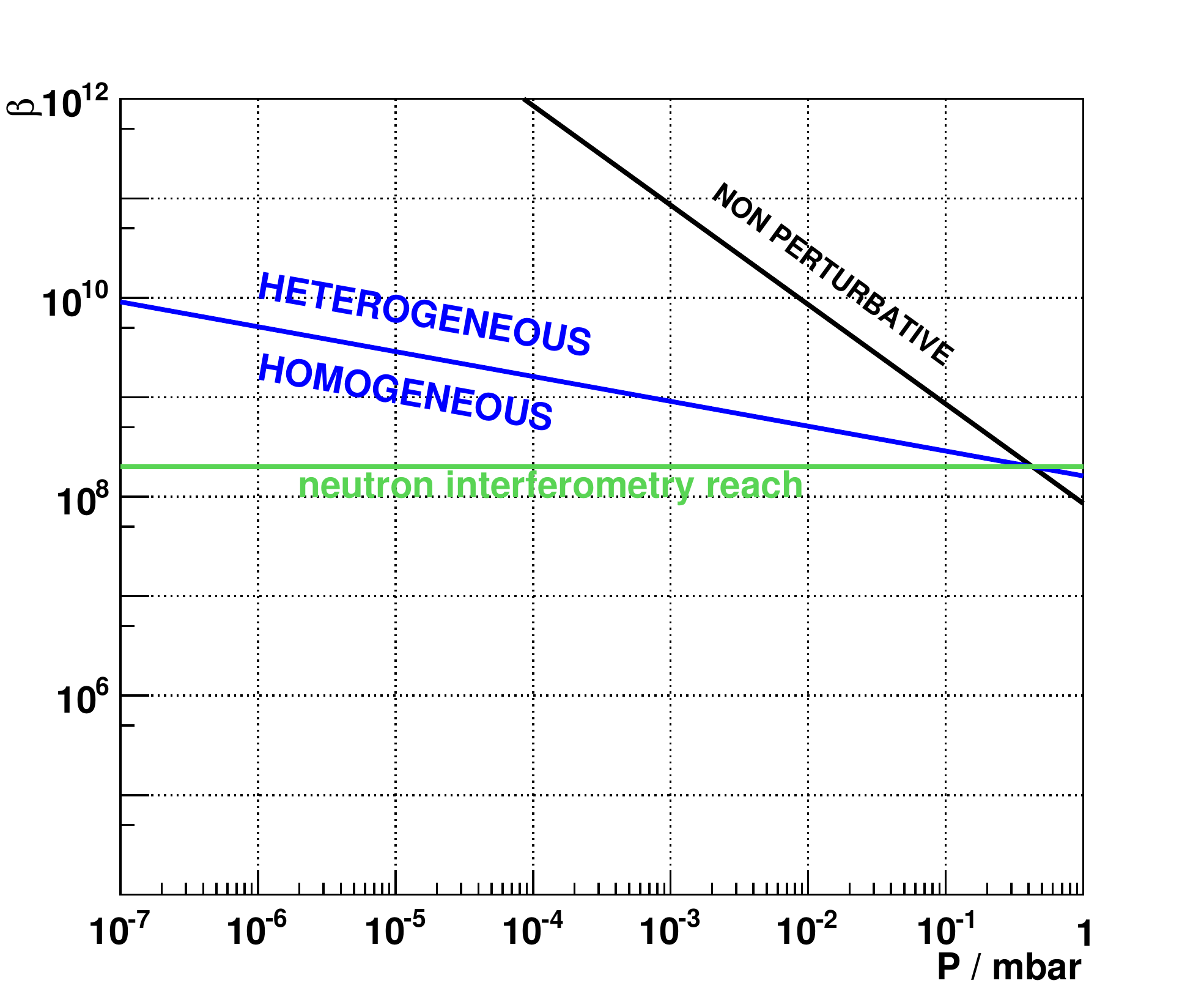}
\caption{The different regimes of the chameleon model in a gas (here helium). For low enough couplings and pressure, the gas can be considered as homogeneous. Conversely for large pressures and couplings, the gas becomes
heterogeneous. The validity of the perturbative approximation in the homogeneous case is confirmed at low enough pressure.}

\label{reg}
\end{figure}

The nuclei are not screened provided
\be
\rho\lesssim \rho_{\rm screen}
\ee
where
\be
\rho_{\rm screen}= \frac{n \Lambda^{4+n}}{M_{\rm Pl}^n}\frac{1}{\beta (2\beta \Phi_{\rm nucl})^{n+1}}.
\ee
When this is the case, the chameleon field outside the nucleus is Coulomb-like with
\be
\varphi (r) =\varphi_G -\frac{\beta m_{\rm nucl}}{M_{\rm Pl}} \frac{1}{4\pi r}
\ee
for distances much less than the large range of the chameleon force.
We have represented in fig. \ref{reg} the different regimes as a function of the coupling $\beta$ and the gas pressure.
For low enough pressures, the gas becomes heterogeneous at large coupling and the treatment presented here is valid. On the other hand, for large pressures, typically larger than 1 mbar, the chameleon becomes
largely self-coupled before the gas can be considered to be heterogeneous and new methods beyond the one presented here must be invoked. This is left for future work.

\subsection{Bubble formation}

When the coupling is too large and the nuclei are screened, the field cannot be considered to be homogeneous. It becomes bubble-like between the nuclei with a shape
akin to the 1D profile of chameleons between two plates.  Immediately outside the nucleus the matter density cannot be taken to be the mean field one anymore and it effectively vanishes implying that
\be
\frac{d^2\varphi}{dr^2} +\frac{2}{r} \frac{d\varphi}{dr} =-n\frac{\Lambda^{n+4}}{\varphi^{n+1}}
\ee
in spherical coordinates as long as the influence of the other nuclei cannot be felt.
The solution to this equation can be well approximated  using the two regions
\be
R_{\rm nucl}<r\le R_\star,\ \ \frac{d^2\varphi}{dr^2} +\frac{2}{r} \frac{d\varphi}{dr} =0
\ee
where $R_\star$ will be determined later and
\be
r>R_\star \ \ \frac{d^2\varphi}{dr^2}=-n\frac{\Lambda^{n+4}}{\varphi^{n+1}}
\ee
This last equation is only valid up to $r=D$ corresponding to the average inter atomic distance where the field is required to have a maximum before falling towards the value $\varphi_{\rm nuc}$ at the neighbouring nuclei.

\begin{figure}
\centering
\includegraphics[width=0.87\linewidth,angle=90]{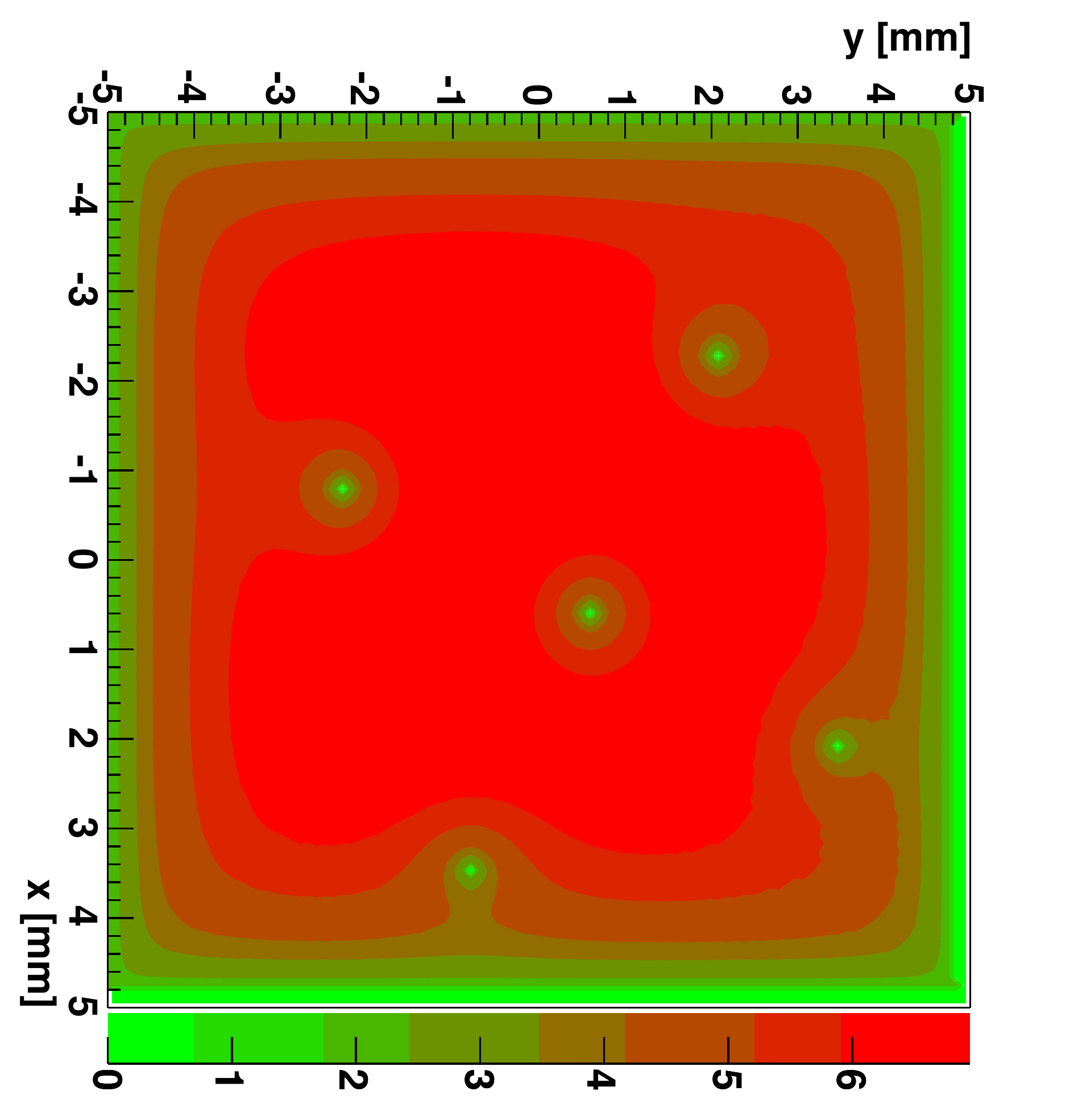}
\caption{The chameleon bubbles in 2D where the field profile is bubble like in a square box and 5 nuclei inside the gas create smaller bubbles.}

\label{hole}
\end{figure}

We find the solution outside the nucleus
\be
R_{\rm nucl} <r\le R_\star \ \ \varphi= B-\frac{C}{r}
\ee
and for $R_\star\le r\le D$
\be
\int_{\varphi_\star}^\varphi\frac{du u^{n/2}}{\sqrt{1-G u^n}} = \sqrt{2} \Lambda^{(n+4)/2} (r-R_\star)
\ee
where $G= \frac{\varphi_G'^2}{2\Lambda^{n+4}}-\frac{1}{\varphi_\star^n}$.
The maximum at $r=D$ is determined by its value
\be
\varphi_D^n= G^{-1}
\ee
implying that
\be
\int_{\varphi_\star/\varphi_D}^1 \frac{ du u^{n/2}}{\sqrt{1-u^n}}= \sqrt 2 \frac{\Lambda^{(4+n)/2}}{\varphi_D^{1+n/2}}(D-R_\star)
\ee
Assuming that $\varphi_\star\ll \varphi_D$ and $D\gg R_\star$ we find that
\be
\varphi_D \approx \left( \frac{\sqrt 2 D \Lambda^{(4+n)/2}}{J_n(0)} \right)^{2/(n+2)}
\ee
Matching at $r=R_{\rm nucl}$ we find that
\be
B= \varphi_{\rm nucl} +\frac{C}{R_{\rm nucl}}
\ee
where $\varphi_{\rm nucl}= \varphi (R_{\rm nucl})$. The field evolves steeply outside the nuclei and therefore
$\varphi_{\rm nucl} \ll  \frac{C}{R_{\rm nucl}}$ implying that
\be
\varphi_\star \approx \frac{C}{R_{\rm nucl}}
\ee
as long as $R_\star \gg R_{\rm nucl}$.
Now $R_\star$ is defined by the condition that
\be
\vert \frac{2}{r}\frac{d\varphi}{dr}\vert_{r=R_\star}= n\frac{\Lambda^{4+n}}{\varphi_\star^{n+1}}
\ee
leading to
\be
R_\star^3 \approx \frac{2 \vert C\vert ^{n+2}}{nR_{\rm nucl}^{n+1} \Lambda^{4+n}}.
\label{rr}
\ee
The solution between $R_{\rm nucl}$ and $R_\star$ is such that the kinetic energy dominates over the potential energy and therefore
$\varphi_\star'^2 \approx \frac{2\Lambda^{4+n}}{\varphi_D^n}$ implying that
\be
C\approx R_\star^2 \sqrt{\frac{2\Lambda^{4+n}}{\varphi_D^n}}.
\ee
Combining
with (\ref{rr}), we find that
the radius $R_\star$ behaves like
\be
R_\star \sim D \left( \frac{R_{\rm nucl}}{D} \right)^{(n+1)/(2n+1)}
\ee
up to irrelevant constants.
When $D\gg R_{\rm nucl}$, we have that $R_\star/R_{\rm nucl} \sim (\frac{R_{\rm nucl}}{D})^{n/(2n+1)}\gg 1$  and $R_\star \ll D$ validating all the approximations.
Hence we have a complete description of the chameleon field outside the nuclei which is independent of the details of the chameleon profile inside the nuclei. On the other hand, it depends
on the typical size of the nuclei $R_{\rm nucl}$. The chameleon profile in the gas comprises bubbles between all the atoms. Each nucleus is surrounded by a small region of radius $R_\star$ wherein the field behaves like a free scalar field. Outside this shell, the field behaves like a 1D bubble whose size is determined by the inter atomic distance.
In particular in the interval between $R_\star$ and $D$, the chameleon profile is universal
\be
\varphi(r) \approx \Lambda (\frac{(2+n)}{\sqrt 2} \Lambda r)^{2/(2+n)}
\ee
valid when $R_\star\ll r\ll D$. For larger distances and therefore well outside the shell of radius $R_\star$ the feel converges to a constant value which depends on the gas pressure.
We have represented in fig. (\ref{hole}) the exact profile of chameleons in 2D when 5 atoms are in a box. We can clearly see that the bubble-like structure emerges.

\subsection{Neutron interferometry and bubbles}

The result on neutron interferometry and the phase of neutrons traversing a chamber where the chameleon has a non-trivial profile between the walls  is only valid as long as the gas does not become a heterogeneous medium for chameleons, something  which happens at large coupling and/or large pressure.
In this case a good approximation for the field profile is that it behaves in a bubble-like fashion between the atoms of the gas. We have seen that the field profile is steep in the interval $R_\star\ll r\ll D$. Apart from this small region around the atoms an therefore between the atoms, the field is nearly constant with a value determined by
\be
\varphi_D \approx (\frac{\sqrt 2 D \Lambda^{(4+n)/2}}{J_n(0)})^{2/(n+2)}
\ee
where $2D$ is the average inter atomic distance. In this regimes we have
\be
\int_{-R}^R \varphi(x) dx \approx 2 R \Lambda (\frac{\sqrt 2 D \Lambda^{(4+n)/2}}{J_n(0)})^{2/(n+2)}
\end{equation}
which is reduced by a power $(D/R)^{2/(n+2)}$ compared to the single bubble case going from one wall of the chamber to the other one. This pre factor follows directly from the
growth rate of bubbles in $r^{2/(n+2)}$, i.e. the bubbles are much smaller as they only grow over a distance $D$ instead of the whole length $R$. As we have obviously  $D\ll R$, the sensitivity of neutron interferometry
in the heterogeneous case becomes extremely poor preventing one to test extremely coupled chameleons. We have represented in fig. (\ref{reg}) the reach of a typical neutron interferometry experiment which lies within the
homogeneous region where the sensitivity is maximal.

\section{Conclusion}

Chameleons are scalar fields which are candidates to model the acceleration of the expansion of the Universe.
Cosmologically, they lead to an equation of state which is close to $w = -1$ and a scalar interaction which would
be detectable by laboratory and solar system tests of gravity if the scalar force were not screened in dense environments.
In this paper we have presented two methods using neutrons which could probe the existence of chameleons experimentally.
The first one involves the chameleon field over a plane mirror and the resulting  perturbation of the neutron energy levels in the terrestrial gravitational field.
The second one uses the quantum properties of neutrons too and interferometry where a sample leading to a chameleon bubble would shift the interference patterns.
In both cases the neutrons are not screened.
In the first case, the neutron wave function in the terrestrial gravitational field is wide enough to prevent screening while in the second case free non-relativistic neutrons have also a large enough Compton wave-length to evade screening. This makes neutrons ideal probes of new screened interactions such as the chameleon one. However we have found that present day sensitivities of both experiments require low enough pressures and large enough couplings of the chameleon to matter, typically $\beta\gtrsim 10^8$. In this regime, it could be that chameleons actually do not see gases as homogeneous media but a large collection of individual atoms (and nuclei). We have analysed this situations and shown that above a certain matter dependent coupling, the chameleon profile in a gas is not homogeneous anymore but bubble-like.
This drastically affects the neutron interferometry sensitivity.
Fortunately, for pressures less than say 1 mbar for helium, neutron interferometry can be sensitive to chameleons with  a coupling less than $\beta\lesssim 10^{10}$ which is within the region of parameter space not already excluded experimentally. As a result, we expect that both the forthcoming GRANIT experiment for bouncing neutrons and the newly designed interferometry experiments will give us interesting results on chameleons at strong coupling.

\bibliographystyle{unsrt}

\end{document}